\documentclass[showpacs,twocolumn,floatfix,superscriptaddress]{revtex4}
\usepackage{graphicx}
\usepackage{amsmath}
\usepackage{amssymb}

\newcommand{\bq}{\begin{equation}}
\newcommand{\ee}{\end{equation}}
\newcommand{\fr}[2]{\frac{#1}{#2}}
\newcommand{\eps}{\varepsilon}

\begin{document}

\title{Pure spin current in graphene NS structures}

\author{D. Greenbaum}
\affiliation{Department of Condensed Matter Physics, Weizmann
Institute of Science, 76100 Rehovot, Israel}
\author{S. Das}
\affiliation{Department of Condensed Matter Physics, Weizmann
Institute of Science, 76100 Rehovot, Israel}
\author{G. Schwiete}
\affiliation{Department of Condensed Matter Physics, Weizmann
Institute of Science, 76100 Rehovot, Israel}
\author{P. G. Silvestrov}
\affiliation{Theoretische Physik III, Ruhr-Universit{\"a}t Bochum,
44780 Bochum, Germany}

\date{\today}

\begin{abstract}
We demonstrate theoretically the possibility of producing a pure
spin current in graphene by filtering the charge from a
spin-polarized electric current. To achieve this effect, which is
based on the recently predicted property of specular Andreev
reflection in graphene, we propose two possible device structures
containing normal-superconductor (NS) junctions.
\end{abstract}



\maketitle

\section{Introduction}

The recent experimental realization of conducting two dimensional
monolayers of
graphite~\cite{Novoselov04,Novosel05,Zhang05,Berger06}, also known
as graphene, offers the promise for new electronic devices. One
conceivable use for graphene is in spintronics~\cite{Wolf01},
where the lack of nuclear spin interaction ($^{12}$C has no
nuclear spin) could offer the ability to maintain spin coherence
over larger distances than in conventional semiconductors. For
progress towards this goal, it is essential to have simple and
reliable means to transport spin in graphene.

We address this issue by proposing a prescription for producing a
pure spin current in ballistic bulk graphene. Spin currents have
already been predicted to arise in graphene due to spin-orbit
coupling~\cite{KaneMele05} and the Quantum Hall
Effect~\cite{Abanin06}. In both cases, the spin currents are due to
counter-propagating edge states of opposite spin. Our proposal makes
use of the recently predicted specularity~\cite{Beenakker06} of
Andreev reflection~\cite{And64} in graphene to produce a pure spin
current in bulk. This is accomplished using structures containing
normal-superconducting (NS) boundaries, which filter the charge out
of a current of spin-polarized quasiparticles, leaving behind a pure
spin current. To be specific, we consider two device paradigms: 1) a
V-junction geometry with opening angle appropriately tuned, 2) a
channel with a superconductor at one boundary and a normal edge at
the other. An advantage of such devices is the large number of
transmitting channels in the bulk, offering the possibility of rapid
spin accumulation. Also, our proposal does not require a magnetic
field, nor does it rely on the spin-orbit gap. In the present case,
however, it is crucial to first generate a spin-polarized electric
current, which could conceivably be done by contacting the system to
a ferromagnetic lead, as has been done with carbon
nanotubes~\cite{ntfm}.

Our description of proximity effects in graphene follows that of
Ref.~\cite{Beenakker06}. Later publications,
Refs.~\cite{Titov06,Bhatta06}, used this approach to discuss the
Andreev spectrum and Josephson effect in NS (SNS) structures in
graphene. A very recent paper~\cite{Titov06Sept} makes explicit
use of the specularity of Andreev reflection in graphene by
considering the neutral excitations propagating along a narrow SNS
channel. These authors propose a device that is similar to the one
considered below in Sec.~\ref{secNSchanel}, but suggest
investigating the thermoelectric effect to observe the chargeless
excitations. In contrast, we analyze the spin transport. Our
approach also lends itself to an analysis of the deviations from
perfect charge filtering in a channel, which is done in
Sec.~\ref{secImperfect}. Finally, an elegant method to produce
pure spin currents in conventional semiconductors was suggested in
Ref.~\cite{Chtchelkatchev}, where the spatial separation of
electron and hole trajectories is caused by tunneling through a
superconductor.

\begin{figure}
\includegraphics[width=5.5cm]{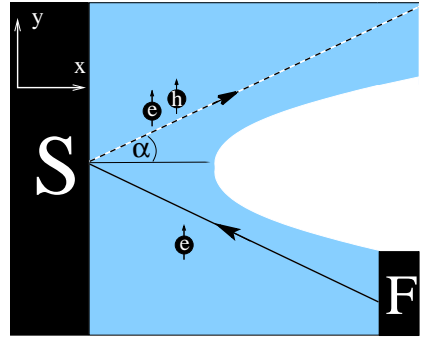}
\caption{ V-junction with NS interface. Polarized electrons are
injected through the lower arm and the same polarization is
transferred to the upper arm by the electron-hole
beam.}\label{vjunct}
\end{figure}

\section{Andreev reflection in graphene}

The electron wave function in graphene is described by a two
component (pseudo-)spinor $\psi$. Its spin-up and spin-down
components correspond to the quantum mechanical amplitudes of
finding the particle on one of the two sublattices of the
honeycomb lattice. The low energy physics of graphene is governed
by two so-called Dirac points in the spectrum, located at the two
inequivalent corners $\vec{K},\vec{K^{\prime }}$ of the Brillouin
zone. The spinor wave function for low energy excitations in
(lightly-doped) graphene decomposes into a sum of two waves
oscillating with different wave vectors $\psi
=e^{i\vec{K}\vec{r}}\phi_+ +e^{i\vec{K'}\vec{r}}\phi_-$. The
smooth envelope functions $\phi_\pm$ satisfy the two-dimensional
Dirac equation~\cite{DiVinMele84} described by the Hamiltonian
 \bq \label{HamGraph}
H_\pm= c(\sigma_x p_x\pm \sigma_y p_y)+U(x),
 \ee
where $c$ is the Fermi velocity and $\vec{p}=-i\hbar\vec{\nabla}$.
In regions of constant $U$ this equation defines a conical energy
band, or {\em valley}, $\eps-U=\pm c|p|$. The Pauli matrices
$\sigma_{x,y}$ permute electrons between two triangular
sublattices of the honeycomb lattice. The two signs in
Eq.~(\ref{HamGraph}) $(+)$ and $(-)$ correspond to the two valleys
$\vec{K}$ and $\vec{K^{\prime }}$.

We consider a graphene sheet in the $x-y$ plane, with the region
$x<0$ covered by a conventional superconductor (Fig.
\ref{vjunct}). Following Ref. \cite{Beenakker06} we assume that in
this configuration a pair potential $\Delta(x)=\Delta_0\Theta(-x)$
can be induced~\cite{Volkov95} in the graphene sheet by the
proximity effect, accompanied by a sufficiently strong shift in
the scalar potential $U(x)=-U_0\Theta(-x)$, so that $U_0\gg
\Delta_0$. The reflection at the NS interface ($x=0$) is described
by a separate {\em four-dimensional} Dirac-Bogoliubov-de~Gennes
equation~\cite{Beenakker06} for each valley
 \bq\label{DBdG}
\left( \begin{array}{cc} H_{\pm}-E_F & \Delta(x) \\
\Delta(x) & E_F-H_{\pm}
\end{array} \right)
\left( \begin{array}{cc} u \\ v
\end{array} \right)=
\eps \left( \begin{array}{cc} u \\ v
\end{array} \right),
 \ee
where the two spinors $u$ and $v$ represent the electron and hole
components of the $\phi_{\pm}$ wavefunctions for $H_{\pm}$,
respectively.

In addition to carrying pseudo-spin, electron-hole, and valley
indices, the quasiparticle wave function should also describe the
usual spin. Since for $\eps<\Delta_0$ no spin can be injected into
the superconductor, the spin of the incident electron is
transferred to the reflected particle.

For several decades it was considered a basic feature of Andreev
reflection~\cite{And64} that the hole produced upon electron-hole
conversion retraces the incident electron's trajectory. Even in
the traditional materials, however, this repetition of
trajectories is exact only for zero excitation energy~$\eps$ in
Eq.\ (\ref{DBdG}). At finite excitation energy the two
trajectories do not quite coincide, leading to interesting effects
in semiclassical dynamics~\cite{Kos95} and the
spectrum~\cite{Sil0306} of Andreev billiards. In graphene, the
Andreev reflection described by Eq.~(\ref{DBdG}) has the standard
form only if $\eps\ll E_F$. Because of the possibility to tune the
Fermi energy close to the Dirac point, one may reach in graphene
the regime $\eps\gg E_F$, so that an incident electron in the
conduction band produces an Andreev reflected hole in the valence
band (opposite side of the cone in the Dirac excitation spectrum).
This valence band hole has the same velocity along the NS
interface as the incident electron, and consequently is specularly
reflected~\cite{Beenakker06}. For vanishing Fermi energy, both the
electron and hole components of the reflected wave follow
precisely the same trajectory. We now restrict ourselves to this
most interesting case of $E_F = 0$.

The transformation of a quantum superposition of incident electron
and hole into the corresponding superposition of outgoing waves
upon Andreev reflection is described by a $2\times 2$ reflection
matrix, $R_A$,
\begin{equation}\label{Rdef}
\left(\begin{array}{c}u\\v\end{array}\right)_{\rm out} =
R_A\left(\begin{array}{c}u\\v\end{array}\right)_{\rm in}=
\left( \begin{array}{cc} r & r_A \\
r_A & r
\end{array} \right)\left(\begin{array}{c}u\\v\end{array}\right)_{\rm in}.
\end{equation}
With the notation $\alpha=\arctan |p_y/p_x|$,
$\eps=c\sqrt{p_x^2+p_y^2}$, and $\xi=\sqrt{\Delta_0^2-\eps^2}$ one
has~\cite{basis}
 \bq\label{R}
r = \frac{-i\eps\sin\alpha}{\eps + i\xi\cos\alpha} \ , \ r_A =
\frac{\Delta_0\cos\alpha}{\eps + i\xi\cos\alpha} \ .
 \ee

\section{ NS interface in V-junction geometry}

Characteristic of specular Andreev reflection in graphene is the
spatial separation of incident and reflected electron-hole beams.
The simplest device that makes use of this property is a V-junction,
as shown in Fig.~1. Suppose one can inject a spin-polarized
collimated monoenergetic beam of electrons through one arm of the
junction. [Monoenergetic beams may potentially be produced by
resonant transport through a graphene quantum dot
(QD)~\cite{SilEf06}, as was done for semiconductor QD's e.g. in
Ref.~\cite{Hohls06}.] The injected electron will be either normally
or Andreev reflected at the NS interface and the created
quasiparticle will escape through the second arm. Since no spin can
penetrate through the superconductor, all the polarization of the
injected beam is transferred to the second arm. On the other hand,
the average charge of the quasiparticles reflected towards the
second arm depends on the excitation energy. One obtains zero
reflected total charge by setting $|r| = |r_A|$. This constrains the
incident particle energy to be
 \bq
\eps=\Delta_0\cot\alpha.
 \ee

Alternatively, one may consider the injection of a collimated beam
of electrons with all possible energies in the range $0<\eps<eV$.
For Dirac particles and a fixed angular spread of the beam,
cancellation of the reflected charge requires
 \bq\label{intint} \int_0^{eV}|r|^2\eps d\eps= \int_0^{eV}
|r_A|^2\eps d\eps.
 \ee
One may think here about electrons leaving the biased~($eV$)
graphene half-plane through a narrow slit and then both angle- and
width-collimated by a second slit. Substitution of Eq.~(\ref{R})
into Eq.~(\ref{intint}) leads to the transcendental equation
 \bq
\ln\left[1+\left(\fr{eV}{\Delta_0}\tan\alpha\right)^2 \right] =
\fr{1}{2}\left(\fr{eV}{\Delta_0}\tan\alpha\right)^2,
 \ee
with the solution
 \bq
eV=1.59 \ \Delta_0 \cot\alpha .
 \ee

Experimental realization of pure spin current in a graphene
V-junction is limited by the requirement of a collimated electron
beam. This difficulty is avoided automatically in the setup proposed
below, where the filtering of charge current takes place because of
multiple Andreev reflections in a NS channel.

\section{ Charge filtering in a long NS channel}\label{secNSchanel}

We consider a normal graphene strip of width $W$ and length $L\gg
W$, which is formed by covering the region $x<0$ of a graphene
half-plane $-\infty< x<W$ with a superconductor (see Fig.~2).
Spin-polarized electrons are injected at $y=0$. We restrict our
discussion to the geometric optics limit, $\lambda \ll W$, where the
transmission through the channel may be considered in terms of
individual particle trajectories. Here $\lambda=2\pi\hbar c/eV$ is
the de~Broglie wavelength for Dirac electrons with energy $eV$. In
addition we consider the low-voltage limit $eV\ll \Delta_0$, since
in this limit the normal reflection at the NS interface is
suppressed~(see Eq.~(\ref{R}))
 \bq\label{lowvoltage}
|r/r_A|^2\sim (eV/\Delta_0)^2\ll 1,
 \ee
and we may assume a perfect electron-hole (hole-electron)
conversion by Andreev reflection.

The number of transmitting channels $N(\eps)$ for the energy
$\eps$ is determined by the width of the strip $W$ and the range
of variation of transverse momentum $|p_x|<\eps/c$,
 \bq
N(\eps)=2\fr{\eps W}{c\hbar \pi}.
 \ee
Here the first factor of $2$ accounts for the two valleys in
graphene. We do not add another $2$ for spin, since we assume
injection of polarized current. [In the limit given in
Eq.~(\ref{lowvoltage}), the possible inter-valley mixing upon
charge- and spin-preserving reflection at the normal edge, see
e.g. Refs.~\cite{Peres06,BreyF06}, is not important for our
calculation.] As usual, each open channel adds $G_0=e^2/h$ to the
differential conductance.

\begin{figure}
\includegraphics[width=8.5cm]{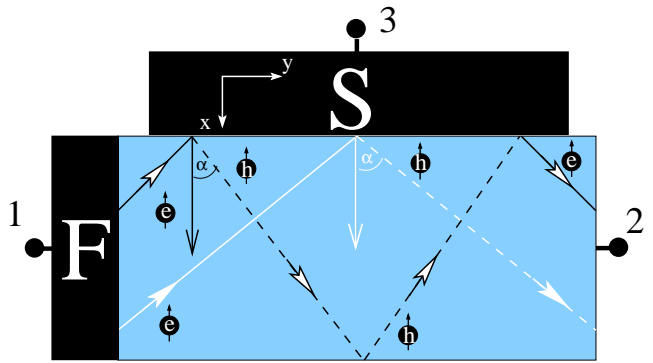}
\caption{ NS-strip. Charge current injected from contact $1$ escapes
to the superconductor $3$, leaving behind a pure spin current
flowing towards electrode $2$.}
\end{figure}

The total current injected into the strip is~\cite{endaccuracy}
 \bq\label{totalcurrent}
I=\int_0^V N(eV')G_0 dV' =\fr{e^2}{h} \fr{eV^2 W}{c\hbar\pi} =
2\fr{e^3V^2 W}{ch^2}.
 \ee
Associated with each incoming electron is a certain quasiparticle
trajectory in the channel as depicted in Fig. 2. In the low-voltage
limit, Eq.~(\ref{lowvoltage}), the outgoing particle is an electron
if the number of reflections from the NS interface is even, and a
hole if the number of reflections is odd. The crucial point is that
for a long channel the contributions to the total current of these
two kinds of trajectories effectively equilibrate, so that the total
charge transfer vanishes after averaging over initial angles. While
the charge current upon such equilibration escapes to the
superconductor, the spin current continues to flow along the strip.
For a $100\%$ polarized injected beam the rate of spin transfer is
simply found as
 \bq
\fr{dS}{dt}=\fr{\hbar}{2}\fr{I}{e}.
 \ee

\begin{figure}
\includegraphics[width=7.5cm]{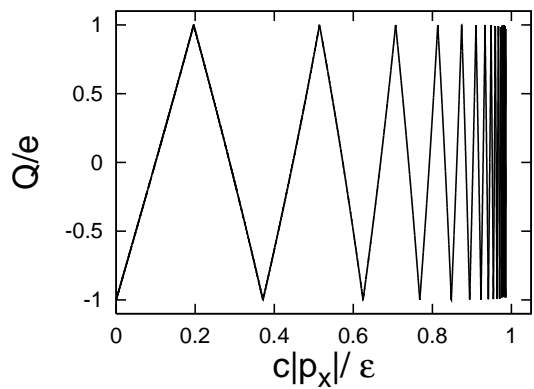}
\caption{ The average charge per quasiparticle after passing
through the strip (length $L$, width $W$, $L/W=10$) as a function
of transverse momentum $p_x$. Very close to $|p_x|=\eps/c$ the
function $Q(p_x)$ could not be shown because of the fast
oscillations.}
\end{figure}

\section{Sources of incomplete charge
filtering}\label{secImperfect}

For a channel of finite length the numbers (weights) of
trajectories with even and odd numbers of Andreev reflections do
not quite coincide, leading to a finite charge current. In this
section we first determine the fraction of charge per
quasiparticle $\langle Q\rangle$ remaining in the beam after
passage through the NS channel, due to this purely geometrical
effect. Since the normal boundary acts as a perfect mirror, we may
effectively consider a strip of doubled width, $0<x<2W$, having
NS-interfaces on both sides. The fluxes of electrons injected into
this doubled strip with
$p_x<0$ and $p_x>0$ are physically equivalent. The charge per
quasiparticle with incident angle $\alpha$ changes linearly along
the channel from $-e$ at $y=0$ to $+e$ at
 \bq
y=y_0=2W\tan\alpha.
 \ee
At $y=y_0$ all electrons injected at angle $\alpha$ are converted
to holes. At $y_0<y<2y_0$ the holes are converted back to
electrons. Charge here changes back (linearly) from $+e$ to $-e$,
and so on.

Figure 3 shows the average charge per quasiparticle for electrons
injected at a fixed angle $\alpha$ at various initial positions,
$0<x<2W$, as a function of $p_x c/\eps=\cos\alpha$ for $L=10W$. The
segments of the function $Q(p_x)$ are given by ($n=0,1,2,\ldots$)
 \bq\label{qcharge}
Q(p_x)=e(-1)^n \left[ \fr{L}{W}\fr{c|p_x|}{\sqrt{\eps^2-(cp_x)^2}}
- (2n+1)\right] .
 \ee
Averaging (integrating) further over $p_x$ and introducing a new
variable $\tau=cp_x/\sqrt{\eps^2-(cp_x)^2}$, we find the charge
fraction $\langle Q\rangle$ per quasiparticle remaining in the
beam to be
 \bq\label{remnantcharge}
\langle Q\rangle =\fr{e}{2}\int_{-\infty}^\infty
F\left(\fr{L}{W}\tau\right)\fr{d\tau}{(1+\tau^2)^{3/2}}.
 \ee
Here we have introduced the sawtooth function $F(x+4)\equiv F(x)$
and $F(x)=|x|-1$ for $|x|<2$. Keeping only the first harmonic in
the Fourier transform of $F(x)$ in the integral in
Eq.~(\ref{remnantcharge}) gives
 \bq\label{remnantfinal}
\langle Q\rangle\approx -\fr{4e}{\pi}\sqrt{\fr{L}{W}} \exp\left(
-\fr{L\pi}{2W} \right).
 \ee
Remarkably, the corrections to charge neutrality are exponentially
small in the limit $L\gg W$. Since this incomplete cancellation of
charge is of geometric origin, Eq.~(\ref{remnantfinal}) does not
depend on the voltage bias $eV$.

{\it Another source of incomplete charge filtering} in the NS
channel is the coexistence of Andreev and normal reflection for
finite quasiparticle energies ($\eps\sim eV\le \Delta_0$). To find
the charge transfer in this case we add to the reflection matrix,
Eq.~(\ref{Rdef}), a part describing particle propagation in
graphene between Andreev reflections
 \bq\label{RdefBig}
R=R_0 R_A.
 \ee
The form of $R_0$ is sensitive to the details of quasiparticle
reflection at the normal side of the strip. Depending on the
microscopic structure of the graphene edge, reflection from it may
or may not introduce transitions between the two valleys
$\vec{K},\vec{K}'$. Here we consider only the latter case as an
example. This also means that $R_0$ is a $2\times 2$ diagonal
matrix.

In the case of decoupled valleys the boundary conditions may only
have the form (see Ref.~\cite{McCannFalko04} for the general
situation)
 \bq\label{boundary}
(a\psi_1 +b\psi_2)|_{x=W}=0\ , \ a,b ={\rm const}
 \ee
where $\psi_1$ and $\psi_2$ are the up and down components of
either the particle ($u$), or hole ($v$) wave function. Particle
number conservation imposes certain restrictions on the allowed
values of $a$ and $b$. Below we consider two examples of such
boundary conditions, demonstrating both existence and absence of
corrections to charge filtering due to finite voltage.

1. The first option is $a/b=\pm i$ (for the edge along the $y$
axis). This boundary is realized if the particle confinement is
achieved by adding a term with large mass $\sigma_z Mc^2$ to the
Dirac equation~(\ref{HamGraph}) at
$x>W$~\cite{BerryMon87,Tworzyd06}. Straightforward calculation for
such a boundary gives
 \bq\label{berymory}
 R_0=ie^{ip_x W/\hbar}{\rm diag}(1,-1) \ , \ (R_0 R_A)^2\propto
{\rm I}.
 \ee
The particle-hole superposition produced upon Andreev reflection
of an electron returns to the pure electron state after the second
reflection from the superconductor. The expectation value of the
quasiparticle charge along a given trajectory according to
Eq.~(\ref{berymory}) switches from $-e$ to $(|r_A|^2-|r|^2)e$ and
back after each reflection from the NS-interface. [The average
charge of the beam with initial angle $\alpha$ now changes
linearly between these two values, not between $-e$ and $e$ as in
Eqs.~(\ref{qcharge},\ref{remnantcharge}).] Averaging over angles
and energy using the exact Eq.~(\ref{R}) gives the charge per
quasiparticle transmitted through the graphene channel for $eV\ll
\Delta_0$ as
 \bq\label{berymoryQ}
\langle Q\rangle = -\fr{\pi}{3}\fr{eV}{\Delta_0}e.
 \ee
We note that according to Eq.~(\ref{R}) the condition $|r/r_A|\ll
1$ is violated for grazing trajectories, having $\pi-\alpha \sim
\eps/\Delta_0$ (\ref{R}). The small statistical weight of these
grazing trajectories is the source of the small value of~$\langle
Q\rangle$.

The result Eq.~(\ref{berymoryQ}) was found in the limit $L\gg W$.
However, further increase of the channel length does not lead here
to charge relaxation, in contrast with the geometric correction,
Eq.~(\ref{remnantfinal}). The possibility for finite charge
current, Eq.~(\ref{berymoryQ}), to flow without leaking along a
(arbitrarily) long NS-interface is very counterintuitive.

2. The second possibility consistent with particle conservation
Eq.~(\ref{boundary}) is $a\times b=0$ (i.e. either $a=0$, or
$b=0$). Boundary conditions of this type describe the bulk
envelope functions in graphene with a zigzag edge~\cite{Abanin06}.
Reflection from the normal edge is now described by a pure phase
matrix $R_0\propto {\rm diag} (1,1)$. Therefore the product of $n$
reflections reduces to
 \bq
(R)^n\propto (R_A)^n\propto{\rm I}\cos n\beta +i\sigma_x\sin
n\beta,\label{R2}
 \ee
where $\cot\beta=-{\eps\tan\alpha}/{\Delta_0}$ and the matrix
$\sigma_x$ interchanges particles and holes. Eq.~(\ref{R2}) leads
to a uniform (to exponential accuracy, as in
Eq.~(\ref{remnantfinal})) mixing of particles and holes after many
reflections at the channel boundaries for any value of $\beta$,
i.e. at any $\eps<\Delta_0$. In contrast to the previous example,
the charge per quasiparticle transmitted through the graphene
channel for $L>>W$ is zero even at finite values of
$\eps/\Delta_0$.

\section{Conclusions}

In summary, we have proposed two basic device concepts for
conversion of polarized electric current into pure spin current in
ballistic bulk graphene. The second device, where the filtering of
charge originates simply from the equilibration of the number of
trajectories experiencing an even or odd number of Andreev
reflections, seems especially promising. We expect this idea to be
easily generalized beyond ballistic transport and for different
geometries. We stress that neither proposal requires a $100\%$
polarization of the incident beam, although the spin transfer rate
will of course depend on the initial polarization.
\\

\section*{Acknowledgements}

DG, SD and GS acknowledge stimulating discussions with Y.~Oreg,
and PGS acknowledges discussions with A.R.~Akhmerov. In addition,
SD thanks V.~Falko and A.~Imamoglu for discussions, and S.~Rao for
useful comments. DG, SD and GS were supported by the Feinberg
Fellowship program at WIS. PGS was supported by the SFB TR 12, and
his visit at Weizmann was supported by the EU - Transnational
Access program, EU project RITA-CT-2003-506095.


\begin{thebibliography}{99}
\bibitem{basis}
The four plane-wave solutions in the normal region used to derive
Eq.~(\ref{R}) are (up to normalization)
 \begin{eqnarray}
(u,v)^\pm_e &=& (\exp(\mp i\alpha/2),\pm\exp(\pm i\alpha/2),0,0),\nonumber\\
(u,v)^\pm_h &=& (0,0,\exp(\mp i\alpha/2),\mp\exp(\pm
i\alpha/2)),\nonumber
 \end{eqnarray}
where $(+)$ denotes $p_x>0$ and $(-)$ denotes $p_x<0$. More
precisely, the $R$-matrix in Eqs.~(\ref{Rdef}) and (\ref{R})
defines the transformation of a linear combination of incoming
$(-)$ electron and hole solutions into a superposition of outgoing
$(+)$ electron and hole solutions.
\bibitem{endaccuracy}
Since the accuracy of the calculation of the number of open
channels, Eq.~(\ref{totalcurrent}), is of order $\delta
N(\eps)\sim 1$, the accuracy of this result is $\delta
I\sim\fr{\lambda}{W}I\sim \fr{e^2}{h}V$. This estimate also
includes the error due to the particles (holes) with energies
close to the Dirac point, $\eps\sim \hbar c/W$, whose dynamics is
not described by the geometric optics approximation.
\end{thebibliography}
\end{document}